\documentclass[10pt,times]{article}
\usepackage{amsmath}
\usepackage{amssymb}
\usepackage{graphicx}
\usepackage{cite}

\usepackage[labelfont=bf,labelsep=period,justification=raggedright]{caption}
\makeatletter
\renewcommand{\@biblabel}[1]{\quad#1.}
\makeatother

\begin{document}
 \begin{flushleft}
{\LARGE
  \textbf{Weak transient chaos}}
\\
\bigskip
{\large
Valentin S. Afraimovich$^1$ and Alexander B. Neiman$^{2}$
}
\\
\bigskip
{\bf 1} IICO-UASLP A. Obreg\'on 64, San Luis Potos\'\i, 78000, M\'exico; 
E-mail: valentin.afraimovich@gmail.com\\
{\bf 2} Department of Physics and Astronomy and Neuroscience Program, Ohio University, Athens, Ohio, USA; E-mail: neimana@ohio.edu\\
\end{flushleft}

\section*{Abstract}
A phenomenon of weak transient chaos is  discussed that is caused by sub-exponential divergence of trajectories in the basin of a non-chaotic attractor. 
Such a regime is not easy to detect, because conventional characteristics, such as the largest Lyapunov exponent is non-positive.
Here we study, how  such a divergence can be exposed and detected.
First, we show that weak transient chaos can be exposed if a small random perturbation is added  to the system, leading to positive values of the largest Lyapunov exponent. 
Second, we introduce an alternative definition of the Lyapunov exponent, which allows us to detect weak transient chaos in the deterministic unperturbed system. 
We show that this novel characteristic becomes positive, reflecting transient chaos.
We demonstrate this phenomenon and its detection using  a master-slave system where the master possesses  a heteroclinic cycle attractor, while the slave is the Van-der-Pol-Duffing oscillator possessing a stable limit cycle.

\section{Introduction}\label{sec:introduction}

Once a new scientific direction arises it influences not only  its immediate area, but also is accompanied by the appearance of new ideas and notions in  neighboring fields. Our article here can be treated as a manifestation of this general principle. About 15 years ago Mikhail Rabinovich with co-authors started considering specific models with the so-called sequential dynamics, based on the winnerless competition principle \cite{RVLHAL} (see also \cite{ATHR,ATVR} and references therein). It turned out that by using such models one can describe and explain important features of  dynamics of neural and cognitive systems.  A collection of works of M. Rabinovich in this direction set up a new area in nonlinear dynamics \cite{MSV,RTV}. Mathematically,  the key point  of these works was the understanding of the important role of heteroclinic networks and heteroclinic channels  for sequential dynamics \cite{AAK,ACN}. 
Trajectories following paths determined by heteroclinic networks may demonstrate behavior that could not be quantitatively described in terms of conventional notions of complexity. Here we describe a regime of weak transient chaos in a model with sequential dynamics and propose a new measure for its quantitative characterization.

There are processes in nature that have adequate mathematical models in the form of dynamical systems (DS) but these models are applicable only for finite intervals of time. For example, a neural network  with parameters determined by a given stimulus behaves  as a specific DS until the instant when another stimulus  arrives. For such systems one should be interested in both  attractors and transient motions. In the seminal works \cite{GOY1,GOY2} C. Grebogi, E. Ott and J. A. Yorke have discovered the phenomenon of transient chaos, explained its  origin, and studied its main features (see \cite{LT} for the  current state of the transient chaos theory). Roughly speaking, if an initial point belongs to the basin of  a regular attractor (e.g. fixed point or a periodic trajectory) and the boundary of the basin contains a chaotic set (e.g. the Smale horseshoe), then the trajectory going through this point behaves chaotically provided that the initial point is close enough to the chaotic set. Of course, if an attractor is chaotic then almost all trajectories in its basin behave chaotically.

Another type of transient motions is observed in systems operating according to the winnerless competition principle. Such motions can be treated as a sequential switching among metastable sets. In the corresponding phase space metastable sets are represented by invariant sets of the saddle type.  For example, saddle equilibrium points or saddle limit cycles may represent metastable sets, while switching is governed by heteroclinic trajectories joining these sets. In this way a heteroclinic network arises and transient motions follow heteroclinic channels around this network's edges. Importantly, such motions  can be chaotic during finite intervals of time \cite{ACY}. However, in contrast to "conventional" transient chaos described by Grebogy-Ott-Yorke \cite{GOY1,GOY2}, this chaotic behavior is not caused by the existence of an unstable chaotic set in the boundary of the basin of an attractor. Hence, this transient dynamics was given the name "finite time chaos" in \cite{ACY}.

The heteroclinic network considered in \cite{ACY} was not an attractor.  That is, each representative point in the heteroclinic channel (that did not belong to the stable manifolds of limit cycles) spent just a finite amount of time inside the channel and then moved away. To make chaotic features of motions more pronounced we consider here the case when a heteroclinic network is an attractor. We consider a master-slave system whereby the master,   represented by $\{x\}$- coordinates, has a stable heteroclinic cycle, an attractor in the $x$-space. Such an attractor consists  of a set of saddle equilibrium points with one-dimensional unstable manifolds;  saddles are joined by heteroclinic trajectories. The slave, represented by $\{y\}$- coordinates, possesses a stable limit cycle in the absence of the master's drive.

\section{Detection of weak transient chaos}
It turns out that the finite time chaos \cite{ACY}  is weaker than usual chaos since the largest Lyapunov exponent vanishes. 
To expose such a behavior one could perturb the master by a weak noise, which would bring its sequential heteroclinic dynamics to a steady state, calculate the Lyapunov exponent of the slave and be convinced in chaos, if the Lyapunov exponent is positive. 
Here we propose an alternative approach: instead of adding noise to the master, we introducing a novel definition of the Lyapunov exponent, which indicates weak transient chaos.

The largest Lyapunov exponent is defined as, 
\begin{equation}
\lambda_T:=\limsup_{t\to\infty}\frac{\ln||D f^t(p)||}{t},
\label{lyap_T}
\end{equation}
where $p$ is an initial point in the basin and $f^t$ is the flow generated by our  master-slave system.  
$\lambda_T=0$, if the divergence of trajectories is subexponential in time. In our situation we have exactly such a case because a trajectory of the master tends to the heteroclinic cycle and spends  progressively more time in neighborhoods of saddle equilibrium points. 
While the representative point is near a saddle  the dynamics is regular: trajectories in the full phase space do not diverge.
The divergence may occur only when the representative point makes transition from one saddle to another.

If one would calculate the topological or the Kolmogorov-Sinai entropy in the case under consideration one obtains 0. The entropy is defined as the limit of a fraction with time $t$ in the denominator, as $t\to\infty$. The numerator measures an amount of instability accumulated in the system up to time $t$. If the numerator grows subexponentially the entropy will be equal to 0. For such situation the so called sequence entropy was introduced by replacing the time $t$ in the denominator by some function of $t$ which increases slower than $t$ \cite{K,Sz,Goo,C}. 
Similar replacement can be probably done for the Lyapunov exponent, resulting in
\begin{equation}
\lambda_{new}:=\limsup_{t\to\infty}\frac{\ln||Df^t(p)||}{\rho(t)}, 
\end{equation} 
instead of (\ref{lyap_T}). However, the disadvantage of this definition is that the denominator $\rho(t)$ will be the same for all initial points $p$, while in reality behavior in time for different $p$ can be different. 

We propose here another approach based on the works of G.M.~Zaslasvky with coauthors \cite{LZ,AZ,EZ,LAB}, reviewed in \cite{L}. It was suggested there to replace (in calculations of complexity functions) time $t$ by the length of a piece of trajectory of temporal length $t$, or some function of it. It was shown that such a replacement allows to obtain an additional useful information in several interesting situations. Following this lead we introduce here a new Lyapunov exponent, 
\begin{equation}
\lambda_S:=\limsup_{t\to\infty}\frac{\ln||Df^t(p)||}{S(p,t)}, 
\label{lyap_S}
\end{equation} 
where $S(p,t)$ is a function of the length of the piece of the trajectory of temporal length $t$ going through an initial point $p$ such that $S\to\infty$ as $t\to\infty$. Let us call $\lambda_S$ the $S$-Lyapunov exponent. It shows how the instability evolves according to the lengths of trajectories.


\section{Master-slave model system}\label{sec:mathematical framework}
We consider a master-slave system in which the master possesses a heteroclinic cycle and drives the slave, Duffing-Van der Pol oscillator.
In the absence of noise the system under consideration has the following form:
\begin{eqnarray}
&&\dot{x}_i=x_i\left(\sigma_i-x_i-\sum_{j\neq i} r_{ij}x_j\right),\ \ i,j=1,2,3, \label{mastsys1}\\
&&\ddot{y}-k\dot{y}(1-y^2)+\alpha y^3+\mu(x)y=0, \label{mastsys2}
\end{eqnarray}
where parameters $\sigma_i, r_{ij}, \alpha, k$ are positive numbers. The master \eqref{mastsys1} drives the slave \eqref{mastsys2} via the coupling function, $\mu(x)$,
\begin{eqnarray}
&&\mu(x)=1+\frac{\varepsilon}{2}\left[1+\tanh\left(z(x)-\Delta\right)\right], \nonumber \\
&&z(x)=[(x_1-x_2)^2+(x_1-x_3)^2+(x_3-x_2)^2],
\label{coupling}
\end{eqnarray}
where $\varepsilon>0$ is the coupling strength and $\Delta>0$ is a threshold parameter. The master system \eqref{mastsys1} has a heteroclinic cycle consisting of three saddle equilibrium points $O_1=(\sigma_1,0,0), O_2=(0,\sigma_2,0)$ and $O_3=(0,0,\sigma_3)$ having one-dimensional unstable manifolds,  connected by  heteroclinic trajectories $\Gamma_{12}, \Gamma_{23}$ and $\Gamma_{31}$.  The following conditions  \cite{AZR},
\begin{eqnarray}
\sigma_2-r_{21}\sigma_1>0,\ \ \sigma_3-r_{31}\sigma_1<0, \label{condition3}\\
\sigma_3-r_{32}\sigma_2>0,\ \ \sigma_1-r_{12}\sigma_2<0, \nonumber \\
\sigma_1-r_{13}\sigma_3>0,\ \ \sigma_2-r_{23}\sigma_3<0, \nonumber
\end{eqnarray} 
guarantee that $O_1, O_2$ and $O_3$ are saddles with one-dimensional unstable manifolds. Furthermore, we assume that 
\begin{eqnarray}
-\sigma_1<\sigma_3-r_{31}\sigma_1, \nonumber \\
-\sigma_2<\sigma_1-r_{12}\sigma_2, \nonumber \\
-\sigma_3<\sigma_2-r_{23}\sigma_3. \nonumber
\end{eqnarray} 
These inequalities imply that the separatrices $\Gamma_{ij}$  approach saddles $O_j$ along the leading direction transversal to the $x_j$- axis, $j=1,2,3.$ Finally, let
\begin{eqnarray}
\nu_1:=-\frac{\sigma_3-r_{31}\sigma_1}{\sigma_2-r_{21}\sigma_1}>1, \nonumber \\
\nu_2:=-\frac{\sigma_1-r_{12}\sigma_2}{\sigma_3-r_{32}\sigma_2}>1, \nonumber \\
\nu_3:=-\frac{\sigma_2-r_{23}\sigma_3}{\sigma_1-r_{13}\sigma_3}>1. \label{condition 11}
\end{eqnarray}
Under these assumptions each saddle $O_j$ is dissipative, and the heteroclinic cycle $\Gamma:=\cup_{j=1}^3O_j\cup(\Gamma_{12}\cup\Gamma_{23}\cup\Gamma_{31})$ 
is an attractor  for the master \eqref{mastsys1}.

We will use the length of the projection onto the $x$-space of the piece of the trajectory of the system \eqref{mastsys1}, \eqref{mastsys2} of  duration $t$ 
 for calculation the length function $S(p, t)$ in the definition of  $S$-Lyapunov exponent, $\lambda_S$. (2). 
Let us remark that the coupling  is effective when the representative point of the master \eqref{mastsys1} is far from the saddles, $O_j$. 
In the full phase space of the system \eqref{mastsys1}, \eqref{mastsys2} there are 3 limit cycles $\{O_i\}\times\{L_i\}$, $i=1,2,3$, where $L_i$ is the limit cycle of the  slave \eqref{mastsys2} for which the coordinates of $O_i$ are substituted into  the coupling function, $\mu(x)$, i.e. $z=2\sigma_i^2$. Each of these cycles is of the saddle type and  have two-dimensional unstable manifolds. While the representative point of the  master  moves from $O_i$ to $O_{i+1}$, the trajectories on the unstable manifold  of the cycle move from one limit cycle $\{O_i\}\times\{L_i\}$ to the next one.  These trajectories thus form a ``heteroclinic tube" that for  vanishing coupling, $\varepsilon=0$ is topologically equivalent (even smoothly equivalent, in fact)  to the direct product $\Gamma\times S'$,  where $S'$ is a circle.  However, as  the coupling strength increases, the  shape of the tube  changes. Its intersection with a section $x_i=a_i>0, a_i\ll 1$, might look as the one on the Fig.~\ref{slave.fig}(b) and so the tube is not a topological manifold anymore.  Such a tube was called {\it a bizarre tube}  in \cite{ACY} and it was shown that a complexity function grows faster for the case of the bizarre tube  compared to piece-wise smooth tubes.



\section{Numerical results}\label{sec:numerical results}
We set the following parameters for the master, $\sigma_1=5, \sigma_2=7, \sigma_3=9,  \Gamma_{12}=1.2243,  \Gamma_{13}=0.0556,  \Gamma_{21}=0.9,  \Gamma_{23}=2.31,  \Gamma_{32}=0.7857$, which satisfy the conditions \eqref{condition3}-\eqref{condition 11} and so the master system possesses a heteroclinic cycle.
The slave Duffing-Van der Pol oscillator \eqref{mastsys2} shows chaotic behavior with positive conventional Lyapunov exponent if perturbed by additive white noise \cite{Goldobin}.
Instead, here it is driven by the master  which possesses a heteroclinic cycle. In the following we set  $k=0.5$, $\alpha=2.5$ for the slave Duffing-Van der Pol oscillator and  $\Delta=8$ for the coupling function, $\mu(x)$. 

We compared deterministic dynamics of the full system with the case when the master  was perturbed by weak additive Gaussian white noise. In this case the master was governed by stochastic differential equations,
\begin{equation}
\dot{x}_i=x_i\left(\sigma_i-x_i-\sum_{j\neq i} r_{ij}x_j\right) + \sqrt{2D}\,\xi_i(t),\ \ i,j=1,2,3, \label{stoch}\\
\end{equation}
where $\xi_i(t)$ are uncorrelated white Gaussian processes and $D$ is their intensity. In the following we set $D=10^{-6}$.

The length of phase trajectories were calculated separately for the master, $S_M$, and for the slave, $S_S$, as
\begin{equation}
S_M  = \int_0^T \left(\sum_{i=1}^3 \dot{x}_i(t)^2\right)^{1/2} dt, \quad
S_S = \int_0^T \left(\dot{y}(t)^2+\ddot{y}(t)^2 \right)^{1/2} dt
\end{equation}
\begin{figure}[h!]
  \centerline{\includegraphics[width=0.9\textwidth]{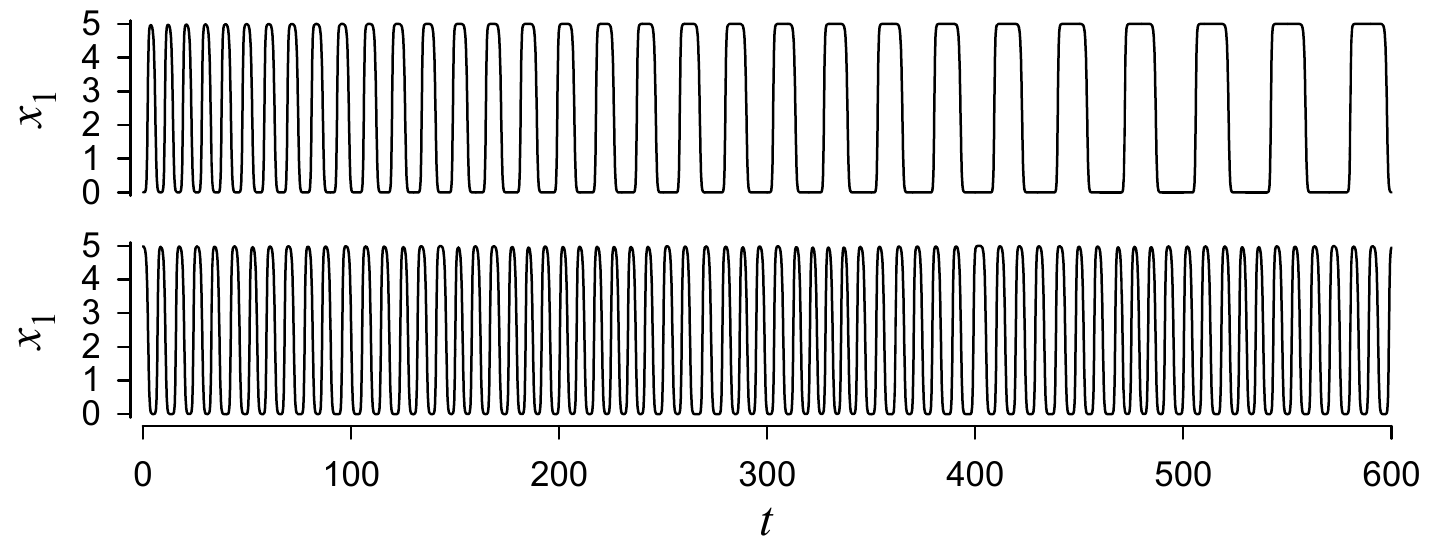}}
  \caption{Time series of the master system. Upper trace: heteroclinic cycle in the intrinsic case. Lower trace: noise perturbed case showing steady state stochastic oscillations.} 
  \label{master.fig}
\end{figure}
Numerical simulations were performed with quadruple precision using 4-th order Runge-Kutta method. To avoid negative values for the master variables we set a reflecting boundary conditions, so that if $x_i(t)<0$, $x_i(t)$ was replaced by $-x_i(t)$.
The largest Lyapunov exponents were calculated for the slave system only over the time span of $10^5$ and additionally averaged over a set of 100 randomly chosen initial conditions of the master and slave systems. We then calculated the standard deviation from this mean, which allows putting "errorbars" on the Lyapunov exponent for indication of its dependence on initial conditions. Both, the conventional Lyapunov exponent, $\lambda_T$ \eqref{lyap_T} and the proposed $S$-exponent with normalization over the length of the projection of the trajectory on master system, $\lambda_S$ \eqref{lyap_S}, were calculated. 
%
%
\begin{figure}[h!]
  \centerline{\includegraphics[width=1.0\textwidth]{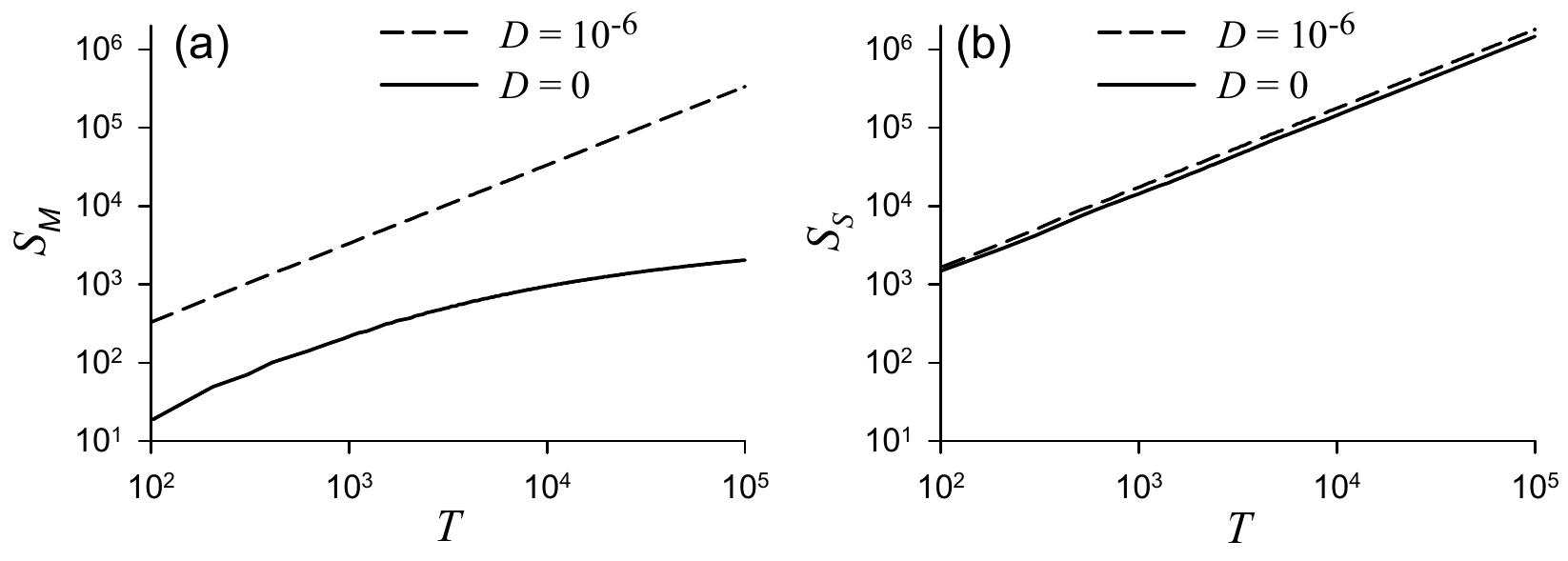}}
  \caption{Trajectory lengths vs integration time. Length of the projection on the master system, $S_M$, (a), and on the slave Duffing-Van der Pol oscillator, $S_S$, (b), are shown for the deterministic case, $D=0$ (solid lines) and for weak noise, $D=10^{-6}$ (dashed line).} 
  \label{length.fig}
\end{figure}

In the absence of noise, $D=0$, the master shows heteroclinic cycle, slowing down as time progresses. That is, the master system generates long transient motions as Fig.\ref{master.fig} (upper trace) indicates. Weak noise accelerates the master  when its phase trajectory passes near saddles and results in steady stochastic oscillations shown in Fig.\ref{master.fig} (lower trace).
As a result, the trajectory length of deterministic slave system shows a limited growth with time, while the length of stochastic slave trajectory exhibit a linear growth, shown in
Fig.\ref{length.fig}(a). From this graph it is easy to see that the average speed of the master (i.e. the slope of $S_M (T)$ on Fig.\ref{length.fig}(a) ) decreases for the deterministic case and is virtually constant for stochastic case. 

\begin{figure}[h!]
  \centerline{\includegraphics[width=1.0\textwidth]{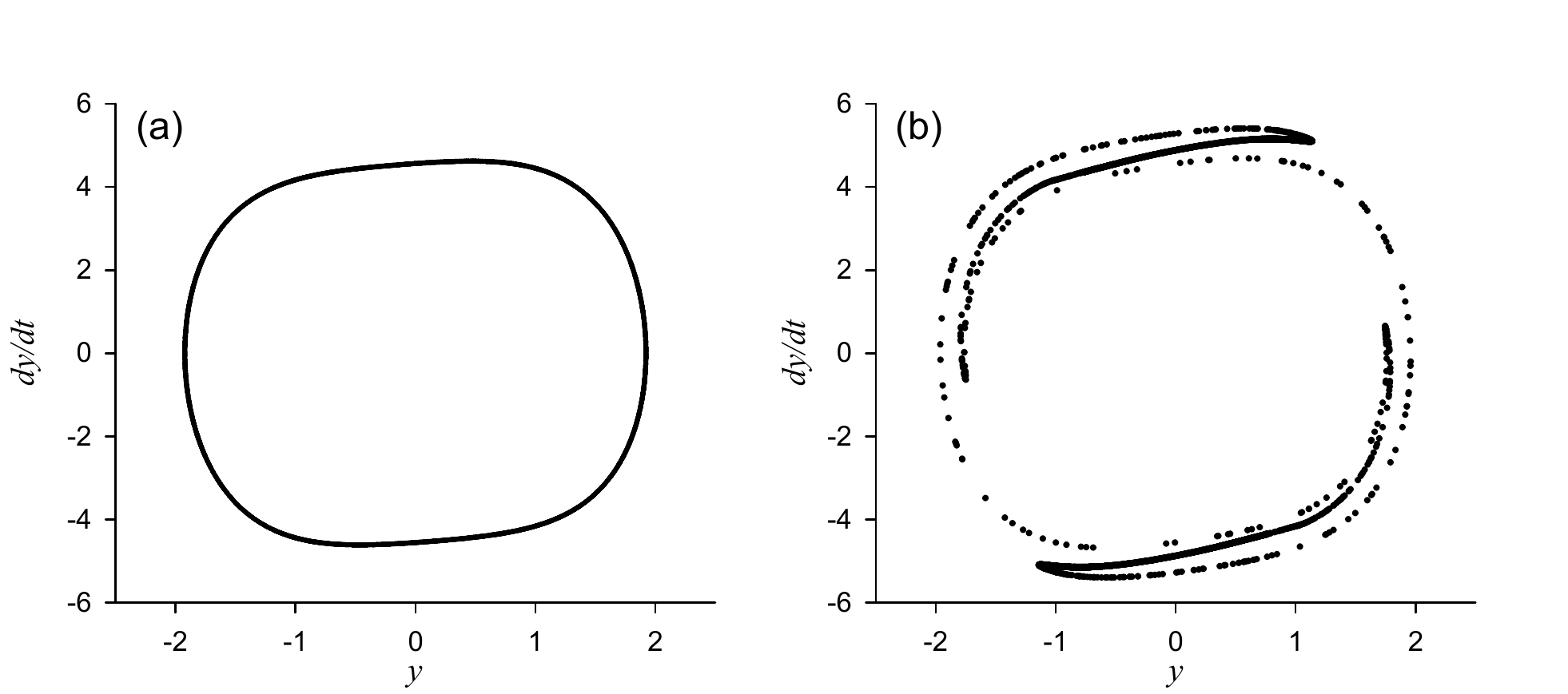}}
  \caption{Effect of drive from heteroclinic cycle of the deterministic master system on the ensemble of 10$^4$ identical Duffing-Van der Pol oscillators. 
  (a): Initial distribution of the ensemble. (b): Distorted distribution after the heteroclinic drive during time $T=10^4$ for $\varepsilon=8$. 
    }
  \label{slave.fig}
\end{figure}
Figure~\ref{slave.fig} shows results of simulations of an ensemble of 10$^4$ identical Duffing-Van der Pol slave oscillators subjected to the common drive from the master system. The ensemble was started with random initial condition, so that before the drive from the master was turned on, the slave oscillators were randomly distributed on the stable limit cycle as Fig.~\ref{slave.fig}(a) shows. Under the influence of the heteroclinic sequence in the master system, the slave's limit cycle can be distorted exhibiting multiple folding as shown in Fig.\ref{slave.fig}(b). 
In both, deterministic and stochastic cases, the slave's trajectory length grows linearly with integration time, $T$, as indicated in Fig.\ref{length.fig}(b). 

\begin{figure}[h!]
  \centerline{\includegraphics[width=1.0\textwidth]{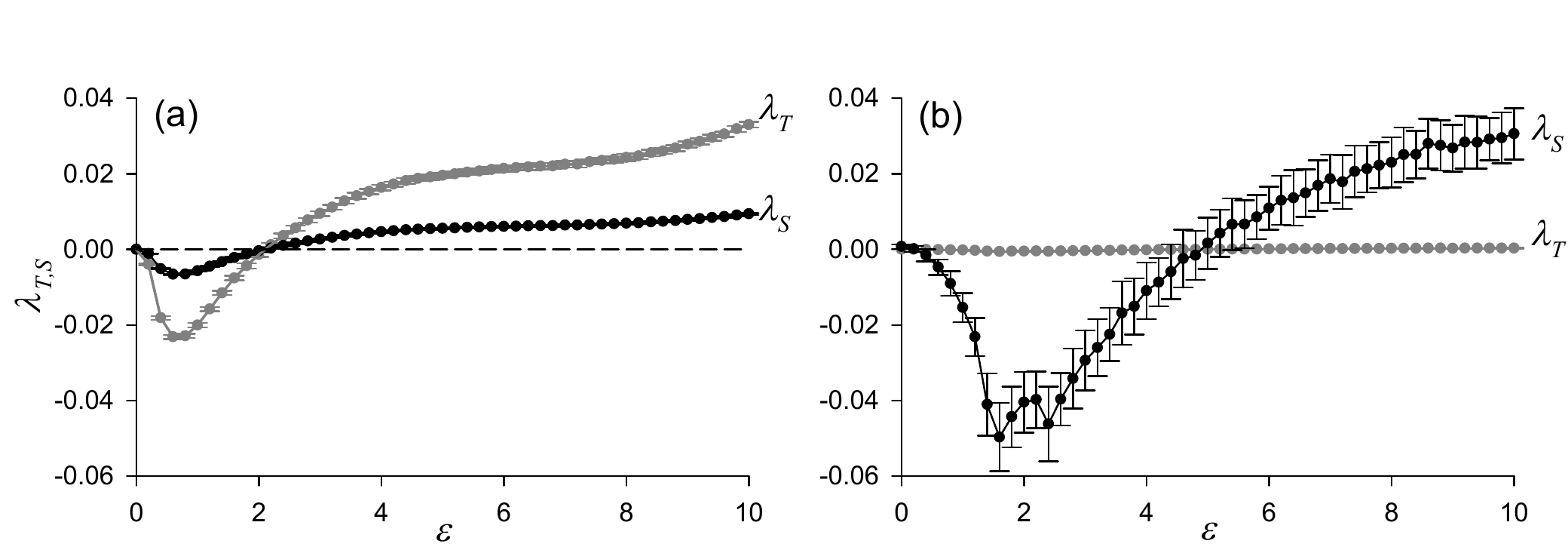}}
  \caption{Largest Lyapunov exponents of the slave system normalized to time, $\lambda_T$ (gray lines and symbols), and to the length of the master system, $\lambda_S$, (black lines and symbols) vs the coupling parameter, $\varepsilon$.  
  Errorbars indicate standard deviation of the Lyapunov exponent from the mean obtained by averaging over 100 random initial conditions.  (a): Randomly perturbed master system, $D=10^{-6}$. (b): Deterministic master system, $D=0$.}
  \label{lyap.fig}
\end{figure}

The results of calculation of the largest Lyapunov exponent are shown in Figure~\ref{lyap.fig}. We begin with the noise-perturbed system which reaches a steady state, Fig.~\ref{lyap.fig}(a). In this case both the conventional and $S$- Lyapunov exponents show qualitatively similar dependence on the coupling parameter, $\varepsilon$: starting with $\varepsilon\approx 2$ both exponents are positive. Noise-induced chaos in this Duffing-Van der Pol oscillator was indeed reported before in \cite{Goldobin}. Importantly, we notice small errorbars, indicating that the Lyapunov exponents are invariant with respect to initial conditions. The deterministic case shown in Fig.\ref{lyap.fig}(b) is different.
The conventional Lyapunov exponent is 0, as expected. However, the $S$ exponent shows positive values for $\varepsilon >5$, indicating transient chaos. Large errorbars point out the dependence on the initial conditions.

\section{Concluding remarks}
We studied weak transient chaos in the master-slave system whereby the slave, possessing  a stable limit cycle, is driven by the master's heteroclinic cycle.
We have shown that if the coupling strength is large enough, then the system manifests a weak transient chaos indicated by positive values of newly introduced $S$-Lyapunov exponent.
We stress that such chaotic behavior is caused neither by the presence of a chaotic unstable set in the boundary of the basin of the attractor (the heteroclinic tube in our case) nor by the chaoticity of the attractor itself. Instead, it is caused by the instability of trajectories in the directions "parallel" to the attractor. Furthermore, on the attractor all trajectories (except for the limit cycles) go from one limit cycle to the next one manifesting trivial non-chaotic behavior. 
Thus, a weak chaotic behavior is caused by divergence of trajectories going through wandering (transient) points. One can say that the chaos is supported on a subset of wandering points. This phenomenon, in slightly different interpretations, was discovered in \cite{Sz} for the sequence entropy.     
Probabilistic distributions of such initial points and/or measures with supports on the set of these points, could not be invariant. From the physical viewpoint these distributions do not correspond to equilibrium or steady-state states. So, in the study of weak transient chaos one should learn how to deal with non-invariant states.


\section*{Acknowledgements} 
The authors dedicate this paper to the 75th anniversary of Mikhail Rabinovich. We wish him great health, and maintenance of his remarkable enthusiasm and energy, which generated and will generate many exciting ideas and fundamental works.
The authors thank T. Young for useful discussions.

\bibliography{references}

\begin{thebibliography}{10}
\providecommand{\url}[1]{\texttt{#1}}
\providecommand{\urlprefix}{URL }
\expandafter\ifx\csname urlstyle\endcsname\relax
  \providecommand{\doi}[1]{doi:\discretionary{}{}{}#1}\else
  \providecommand{\doi}{doi:\discretionary{}{}{}\begingroup
  \urlstyle{rm}\Url}\fi
\providecommand{\bibAnnoteFile}[1]{%
  \IfFileExists{#1}{\begin{quotation}\noindent\textsc{Key:} #1\\
  \textsc{Annotation:}\ \input{#1}\end{quotation}}{}}
\providecommand{\bibAnnote}[2]{%
  \begin{quotation}\noindent\textsc{Key:} #1\\
  \textsc{Annotation:}\ #2\end{quotation}}
\providecommand{\eprint}[2][]{\url{#2}}

\bibitem{RVLHAL}
Rabinovich M, Volkovskii A, Lecanda P, Huerta R, Abarbanel H, et~al. (2001)
  Dynamical encoding by networks of competing neuron groups: winnerless
  competition.
\newblock Physical review letters 87: 068102.
\bibAnnoteFile{RVLHAL}

\bibitem{ATHR}
Afraimovich V, Tristan I, Huerta R, Rabinovich MI (2008) Winnerless competition
  principle and prediction of the transient dynamics in a lotka--volterra
  model.
\newblock Chaos: An Interdisciplinary Journal of Nonlinear Science 18: 043103.
\bibAnnoteFile{ATHR}

\bibitem{ATVR}
Afraimovich V, Tristan I, Varona P, Rabinovich M (2013) Transient dynamics in
  complex systems: Heteroclinic sequences with multidimensional unstable
  manifolds. discontinuity.
\newblock Nonlinearity and Complexity 2: 21--41.
\bibAnnoteFile{ATVR}

\bibitem{MSV}
Rabinovich MI, Simmons AN, Varona P (2015) Dynamical bridge between brain and
  mind.
\newblock Trends in cognitive sciences 19: 453--461.
\bibAnnoteFile{MSV}

\bibitem{RTV}
Rabinovich MI, Tristan I, Varona P (2015) Hierarchical nonlinear dynamics of
  human attention.
\newblock Neuroscience \& Biobehavioral Reviews 55: 18--35.
\bibAnnoteFile{RTV}

\bibitem{AAK}
Afraimovich V, Ashwin P, Kirk V (2010) Robust heteroclinic and switching
  dynamics.
\newblock Dynamical Systems 25: 285--286.
\bibAnnoteFile{AAK}

\bibitem{ACN}
Ashwin P, Coombes S, Nicks R (2016) Mathematical frameworks for oscillatory
  network dynamics in neuroscience.
\newblock The Journal of Mathematical Neuroscience 6: 1--92.
\bibAnnoteFile{ACN}

\bibitem{GOY1}
Grebogi C, Ott E, Yorke JA (1983) Crises, sudden changes in chaotic attractors,
  and transient chaos.
\newblock Physica D: Nonlinear Phenomena 7: 181--200.
\bibAnnoteFile{GOY1}

\bibitem{GOY2}
Grebogi C, Ott E, Yorke JA (1986) Critical exponent of chaotic transients in
  nonlinear dynamical systems.
\newblock Physical review letters 57: 1284.
\bibAnnoteFile{GOY2}

\bibitem{LT}
Lai YC, T{\'e}l T (2011) Transient chaos: complex dynamics on finite time
  scales, volume 173.
\newblock Springer Science \& Business Media.
\bibAnnoteFile{LT}

\bibitem{ACY}
Afraimovich V, Cuevas D, Young T (2013) Sequential dynamics of master--slave
  systems.
\newblock Dynamical Systems 28: 154--172.
\bibAnnoteFile{ACY}

\bibitem{K}
Kushnirenko AG (1967) On metric invariants of entropy type.
\newblock Russian Mathematical Surveys 22: 53--61.
\bibAnnoteFile{K}

\bibitem{Sz}
Szlenk W (1979) On weakly* conditionally compact dynamical systems.
\newblock Studia Mathematica 66: 25--32.
\bibAnnoteFile{Sz}

\bibitem{Goo}
Goodman T (1974) Topological sequence entropy.
\newblock Proc London Math Soc 29: 331--350.
\bibAnnoteFile{Goo}

\bibitem{C}
Canovas J (2008) Progress in Mathematical Biology Research, Nova Science
  Publishers, chapter A Guide to Topological Sequence Entropy.
\newblock pp. 101 - 139.
\bibAnnoteFile{C}

\bibitem{LZ}
Leoncini X, Zaslavsky GM (2002) Jets, stickiness, and anomalous transport.
\newblock Physical Review E 65: 046216.
\bibAnnoteFile{LZ}

\bibitem{AZ}
Afraimovich V, Zaslavsky G (2003) Space--time complexity in hamiltonian
  dynamics.
\newblock Chaos: An Interdisciplinary Journal of Nonlinear Science 13:
  519--532.
\bibAnnoteFile{AZ}

\bibitem{EZ}
Zaslavsky G, Edelman M (2005) Polynomial dispersion of trajectories in sticky
  dynamics.
\newblock Physical Review E 72: 036204.
\bibAnnoteFile{EZ}

\bibitem{LAB}
Leoncini X, Agullo O, Benkadda S, Zaslavsky GM (2005) Anomalous transport in
  charney-hasegawa-mima flows.
\newblock Physical Review E 72: 026218.
\bibAnnoteFile{LAB}

\bibitem{L}
Leoncini X (2010) Hamiltonian chaos and anomalous transport in two dimensional
  flows.
\newblock In: Hamiltonian Chaos Beyond the KAM Theory, Springer. pp. 143--192.
\bibAnnoteFile{L}

\bibitem{AZR}
Afraimovich V, Zhigulin V, Rabinovich M (2004) On the origin of reproducible
  sequential activity in neural circuits.
\newblock Chaos: An Interdisciplinary Journal of Nonlinear Science 14:
  1123--1129.
\bibAnnoteFile{AZR}

\bibitem{Goldobin}
Goldobin DS, Pikovsky A (2005) Synchronization and desynchronization of
  self-sustained oscillators by common noise.
\newblock Physical Review E 71: 045201.
\bibAnnoteFile{Goldobin}

\end{thebibliography}

\end{document}